\documentclass[twocolumn]{aastex62}
\usepackage{ulem}
\hypersetup{linkcolor=red,citecolor=blue,filecolor=cyan,urlcolor=magenta}
\usepackage{amsmath}
\usepackage{subfigure}
\usepackage{float}
\usepackage{tensor}     
\usepackage{graphicx}   
\usepackage{bm}         
\usepackage{xcolor}     
\usepackage{color}      
\usepackage{longtable}  
\usepackage[section]{placeins} 
\usepackage[utf8]{inputenc} 
\newcommand{\nc}{\newcommand*} 
\nc{\figurewidth}{3.2in}
\nc{\xbar}{\bar{x}}
\nc{\rhoeq}{\rho_{\mathrm{eq}}}
\nc{\zeq}{z_{\mathrm{eq}}}
\nc{\tla}{\tilde{\lambda}}
\nc{\dt}{\delta}
\nc{\Dt}{\Delta}
\nc{\vj}{\vec{j}}
\nc{\vl}{\vec{l}}
\nc{\hx}{\hat{x}}
\nc{\hy}{\hat{y}}
\nc{\bj}{\bm{j}}
\nc{\mJ}{\mathcal{J}}
\nc{\mP}{\mathcal{P}}
\nc{\Msun}{M_\odot}
\nc{\app}{\approx}
\nc{\av}[1]{\langle #1 \rangle}
\nc{\eq}[1]{Eq.~\eqref{#1}}
\nc{\al}{\alpha}
\nc{\Xstar}{X_{\ast}}
\nc{\seq}{\sigma_{\mathrm{eq}}}
\nc{\fpbh}{f_{\mathrm{pbh}}}
\nc{\vth}{\vec{\theta}}
\nc{\vla}{\vec{\lambda}}
\nc{\vd}{\vec{d}}
\nc{\Mmin}{M_{\mathrm{min}}}
\nc{\rmd}{\mathrm{d}}
\nc{\mmin}{{m_{\mathrm{min}}}}
\nc{\mmax}{{m_{\mathrm{max}}}}
\nc{\mR}{\mathcal{R}}
\nc{\tmR}{\tilde{\mathcal{R}}}
\nc{\s}{\sigma}
\nc{\ogw}{\Omega_{\mathrm{GW}}}
\nc{\addref}{[\textcolor{red}{add ref}] }
\nc{\Om}{\Omega}
\nc{\gm}{\gamma}
\nc{\Gm}{\Gamma}
\nc{\gpcyr}{\mathrm{Gpc}^{-3}\,\mathrm{yr}^{-1}}
\nc{\Eq}[1]{Eq.~\eqref{#1}}
\nc{\Fig}[1]{Fig.~\ref{#1}}
\nc{\Table}[1]{Table~\ref{#1}}
\nc{\lvc}{LIGO/Virgo} 
\nc{\Sec}[1]{Sec.~\ref{#1}}
\nc{\eg}{\textit{e.g.~}}
\nc{\SNR}{\mathrm{SNR}}
\nc{\bt}{\mathbf{t}}
\nc{\be}{\mathbf{\epsilon}}
\nc{\bn}{\mathbf{n}}
\nc{\bd}{\mathbf{d}}
\nc{\ba}{\mathbf{a}}
\nc{\bnu}{\mathbf{\nu}}
\nc{\uni}{\mathrm{U}}
\nc{\logu}{\operatorname{\mathrm{log-U}}}
\nc{\RN}{\mathrm{RN}}
\nc{\BN}{\mathrm{BN}}
\nc{\GN}{\mathrm{GN}}
\nc{\mcN}{\mathcal{N}}
\nc{\GWB}{\mathrm{GW}}
\nc{\yr}{\mathrm{yr}}
\nc{\Am}{\mathcal{A}}
\nc{\Dm}{\mathcal{D}}
\nc{\Hm}{\mathcal{H}}

\nc{\mrm}{\mathrm}
\nc{\BE}{B\scriptsize{AYES}\normalsize{E}\scriptsize{PHEM}\normalsize  }

\nc{\Ostgw}{\Omega_{\mathrm{GW}}^{\mathrm{ST}}}
\nc{\Ottgw}{\Omega_{\mathrm{GW}}^{\mathrm{TT}}}
\nc{\Ovlgw}{\Omega_{\mathrm{GW}}^{\mathrm{VL}}}
\nc{\Oslgw}{\Omega_{\mathrm{GW}}^{\mathrm{SL}}}

\def\half{{1\over 2}}

\def\half{{1\over 2}}
\def\({\left(}
\def\){\right)}
\def\[{\left[}
\def\]{\right]}

\def\e{\begin{equation}}
	\def\q{\end{equation}}
\def\m{\begin{eqnarray}}
	\def\n{\end{eqnarray}}
\nc{\red}[1]{\textcolor{red}{#1}}
\received{}
\revised{}
\accepted{}
\published{}
\submitjournal{ApJ}
\begin{document}
	
\title{Constraining the Polarization of Gravitational Waves with the Parkes Pulsar Timing Array Second Data Release}       

\author{Yu-Mei Wu}
\email{wuyumei@itp.ac.cn} 
\affiliation{CAS Key Laboratory of Theoretical Physics, Institute of Theoretical Physics, Chinese Academy of Sciences, Beijing 100190, China}
\affiliation{School of Physical Sciences, University of Chinese Academy of Sciences, No. 19A Yuquan Road, Beijing 100049, China}

\author{Zu-Cheng Chen}
\email{Corresponding author: chenzucheng@itp.ac.cn} 
\affiliation{CAS Key Laboratory of Theoretical Physics, Institute of Theoretical Physics, Chinese Academy of Sciences, Beijing 100190, China}
\affiliation{School of Physical Sciences, University of Chinese Academy of Sciences, No. 19A Yuquan Road, Beijing 100049, China}

\author{Qing-Guo Huang}
\email{Corresponding author: huangqg@itp.ac.cn}
\affiliation{CAS Key Laboratory of Theoretical Physics, 
    Institute of Theoretical Physics, Chinese Academy of Sciences,
    Beijing 100190, China}
\affiliation{School of Physical Sciences, 
    University of Chinese Academy of Sciences, 
    No. 19A Yuquan Road, Beijing 100049, China}
\affiliation{School of Fundamental Physics and Mathematical Sciences,
    Hangzhou Institute for Advanced Study, UCAS, Hangzhou 310024, China}
\affiliation{Center for Gravitation and Cosmology, 
    College of Physical Science and Technology, 
    Yangzhou University, Yangzhou 225009, China}

\begin{abstract}
We search for the isotropic stochastic gravitational-wave background, including the nontensorial polarizations that are allowed in general metric theories of gravity, in the Parkes Pulsar Timing Array (PPTA) second data release (DR2). We find no statistically significant evidence that the common-spectrum process reported by the PPTA collaboration has the tensor transverse, scalar transverse, vector longitudinal, or scalar longitudinal correlations in PPTA DR2. Therefore, we place a $95\%$ upper limit on the amplitude of each polarization mode, as $\mathcal{A}_{\mathrm{TT}} \lesssim 3.2\times 10^{-15}$, $\mathcal{A}_{\mathrm{ST}} \lesssim 1.8\times 10^{-15}$, $\mathcal{A}_{\mathrm{VL}}\lesssim 3.5\times 10^{-16}$ and $\mathcal{A}_{\mathrm{SL}}\lesssim 4.2\times 10^{-17}$; or, equivalently, a $95\%$ upper limit on the energy density parameter per logarithm frequency, as $\Omega_{\mathrm{GW}}^{\mathrm{TT}} \lesssim 1.4\times 10^{-8}$, $\Omega_{\mathrm{GW}}^{\mathrm{ST}} \lesssim 4.5\times 10^{-9}$, $\Omega_{\mathrm{GW}}^{\mathrm{VL}} \lesssim 1.7\times 10^{-10}$ and $\Omega_{\mathrm{GW}}^{\mathrm{SL}} \lesssim 2.4\times 10^{-12}$, at a frequency of 1/yr.
		
\end{abstract}
	
	
\section{Introduction}

The thrilling direct detection of the gravitational waves (GWs) from a binary black hole coalescence \citep{LIGOScientific:2016aoc} validates the feasibility of utilizing GWs as a gravitational test tool and marks a new era for gravitational astronomy. Different kinds of GW detectors are sensitive to different frequency bands. For instance, the current ground-based detectors are sensitive to the $10 \sim 10^4$\,Hz band \citep{Abbott:2016xvh}, while the future space-based detectors will govern the regime of $10^{-5} \sim 1$\,Hz \citep{Gair:2012nm}. For the GWs in a much lower frequency band, the nanohertz band, Pulsar Timing Arrays (PTAs) are natural detectors, as they measure the times of arrival (TOAs) of radio pulses from stable millisecond pulsars over timescales of years \citep{1978SvA....22...36S,Detweiler:1979wn,1990ApJ...361..300F}. Specifically, millisecond pulsars emit radio pulses at extremely stable rates, and the arrival of such pulses at the Earth can be timed with high precision, making the pulsars sensitive probes of their environment. When there is a difference between the expected arrival time, as described by the timing model that characterizes the pulsar's spin period, proper motion, and binary orbital dynamics, etc., and the actual arrival time, this indicates that unmodeled effects exist within the data, including those caused by GWs.  Furthermore, if a collection of pulsars is exposed in the space-time disturbed by GWs, the signals will be encoded as spatially correlated fluctuations, which discriminate the GW signals of interest from other sources of noise.  As the observational time and the number of pulsars increase, the sensitivity of the PTA gets better. Currently, three major PTA collaborations are involved in the effort of searching for GWs in the nanohertz frequency band, including the North American Nanohertz Observatory for GWs (NANOGrav) \citep{McLaughlin:2013ira}, the European Pulsar Timing Array (EPTA) \citep{Kramer:2013kea} and the Parkes Pulsar Timing Array (PPTA) \citep{Manchester:2012za}. These three PTAs collaborate as the International Pulsar Timing Array (IPTA) \citep{Hobbs:2009yy,Manchester:2013ndt}.

The stochastic gravitational-wave background (SGWB)--the primary goal of the search of the PTA collaborations--is expected to be dominant in nanohertz band, which might originate from the supermassive black hole binaries (SMBHBs) \citep{1995ApJ...446..543R,Sesana:2012ak}, comic strings \citep{Damour:2004kw,Blanco-Pillado:2017rnf}, the first phase transition \citep{Caprini:2010xv}, and scalar-induced GWs \citep{Yuan:2019udt}. Over the last few decades, the PTA collaborations have not found the GW signals, but the increasingly sensitive data sets offer increasingly stringent constraints on the SGWB \citep{vanHaasteren:2011ni,Lentati:2015qwp,NANOGrav:2015aud,NANOGRAV:2018hou,2013Sci...342..334S,2015Sci...349.1522S,Chen:2019xse}. Recently, the NANOGrav collaboration reported that there is strong evidence in favor of a stochastic common-spectrum process, which is modeled by a power-law spectrum among the pulsars, over the independent red noise processes of each pulsar, in their 12.5 yr data set \citep{Arzoumanian:2020vkk}. However, given the lack of statistically significant evidence for the quadrupolar spatial correlations, it is inconclusive to claim the detection of an SGWB consistent with general relativity. Note that the tensor transverse (TT) modes giving rise to the quadrupolar spatial correlations constitute only two of the six GW polarization modes that are allowed in a general metric theory of gravity, which also includes one scalar transverse (ST) mode, two vectorial longitudinal (VL) modes, and one scalar longitudinal (SL) mode. Later on, \cite{Chen:2021wdo} searched for nontensorial SGWBs in the NANOGrav 12.5 yr data set, and found strong Bayesian evidence that the common-spectrum reported by the NANOGrav collaboration had ST spatial correlations. More recently, the PPTA collaboration has also found a common-spectrum process in their second data release (DR2), with the PPTA DR2 showing no significant evidence for, or against, the TT spatial correlations \citep{Goncharov:2021oub}. 

In this article, we mainly search for the nontensorial polarizations that are allowed in a general metric theory of gravity in the PPTA DR2 \citep{Kerr:2020qdo}, which comprises the observations of 26 pulsars with time spans as long as 15 yr. We find no statistically significant evidence that the common-spectrum process reported by the PPTA collaboration has TT, ST, VL, or SL spatial correlations. Therefore, we place upper limits on the amplitudes of the corresponding polarization modes.

\section{SGWB from a general metric theory}\label{sec2}

The superposition of numerous unresolved GW signals from a cosmic uniformly distributed population of SMBHBs results in an isotropic SGWB. Assuming the binaries are inspiraling in circular orbits, and their orbital evolutions are dominated by GW emissions, the characteristic GW strain can be modeled as a power-law spectrum \citep{Sampson:2015ada, Cornish:2017oic}
\m
    h_c(f)= A_{\GWB}\left(\frac{f}{f_{\yr}}\right)^\alpha,	
\n
where $A_{\GWB}$ is the amplitude of the strain spectrum of the SGWB measured at $f_{\yr}=1/\rm{yr}$ and $\alpha$ is a power index that takes a value of $\alpha=-2/3$ for the TT mode, which is dominated by quadrupole radiation, and $\alpha=-1$ for all of the ST, VL, and SL modes, which are dominated by dipole radiation \citep{Cornish:2017oic, Sampson:2015ada}. The dimensionless GW energy density parameter per logarithm frequency is related to the amplitude $A_{\GWB}$ by \cite{Thrane:2013oya}:
\m
    \Om_{\rm{GW}}(f) = \frac{2\pi^2 }{3H_0^2}h_c^2 f^2=\frac{2\pi^2A_{\GWB}^2}{3H_0^2}\(\frac{f}{f_{\yr}}\)^{2+2\alpha}f_{\yr}^2,
\n
with the Hubble constant $H_0=67.4\, \rm{km\,s^{-1}}$ $\rm{Mpc^{-1}}$ from Planck 2018 \citep{Planck:2018vyg}.

\begin{table*}[!htbp]
    \caption{The Noise Models for the 25 Pulsars Used in this Work from PPTA DR2.}
    \label{noise}
    \begin{tabular}{c|c c c|c c c |c c|c c c}
        \hline
        Pulsar &EFAC& EQUAD &ECORR & SN &DM & CN$^{\rm a}$& BN &GN& $E_{dip}$$^{\rm b}$& $G_{bump}$$^{\rm c}$ & $A_{dm}$$^{\rm d}$\,\\
        \hline
        J0613-0200\, & $\checkmark$& $\checkmark$ & $\checkmark$& $\checkmark$ & &4& & &    & &$\checkmark$\, \\
        \hline
        J0711-6830&$\checkmark$ &$\checkmark$& & $\checkmark$&$\checkmark$& &&&&&\\
        \hline
        J1017-7156&$\checkmark$ &$\checkmark$&$\checkmark$& & & 2.29& & & & &\\
        \hline
        J1022+1001&$\checkmark$ &$\checkmark$&$\checkmark$& &$\checkmark$& & &&&&\\
        \hline
        J1024-0719&$\checkmark$ &$\checkmark$& & &$\checkmark$& & &&&&\\
        \hline
        J1045-4509&$\checkmark$ &$\checkmark$& & && 1.82& &&&&\\
        \hline
        J1125-6014&$\checkmark$ &$\checkmark$& & &$\checkmark$& & &&&&\\
        \hline
        J1446-4701&$\checkmark$ &$\checkmark$&$\checkmark$ & &$\checkmark$& & &&&&\\
        \hline
        J1545-4550&$\checkmark$ &$\checkmark$& & &$\checkmark$& & &&&&\\
        \hline
        J1600-3053&$\checkmark$ &$\checkmark$&$\checkmark$ & $\checkmark$&$\checkmark$& & \small{-B 40CM \,-B 50CM}&&&&\\
        \hline
        J1603-7202&$\checkmark$ &$\checkmark$&$\checkmark$ & & $\checkmark$& & &&&$\checkmark$&\\
        \hline
        J1643-1224&$\checkmark$ &$\checkmark$& & $\checkmark$&$\checkmark$& & \small{-B 40CM \,-B 50CM}&&$\checkmark$&&\\
        \hline
        J1713+0747&$\checkmark$ &$\checkmark$& $\checkmark$& &$\checkmark$& & \small{-B 10CM \,-B 20CM} & \small{CPSR2\_20CM} &$\checkmark$&&\\
        \hline
        J1730-2304&$\checkmark$ &$\checkmark$& & &$\checkmark$& & &&&&\\
        \hline
        J1732-5049&$\checkmark$ &$\checkmark$& & &$\checkmark$& & &&&&\\
        \hline
        J1744-1134&$\checkmark$ &$\checkmark$&$\checkmark$ & &$\checkmark$& & \small{-B 10CM \,-B 20CM}&&&&\\
        \hline
        J1824-2452A&$\checkmark$ &$\checkmark$&$\checkmark$ & $\checkmark$&$\checkmark$& &\small{-B 40CM \,-B 50CM}&&&&\\
        \hline
        J1832-0836&$\checkmark$ &$\checkmark$& & &$\checkmark$& & &&&&\\
        \hline
        J1857+0943&$\checkmark$ &$\checkmark$& & $\checkmark$&$\checkmark$& & &&&&\\
        \hline
        J1909-3744&$\checkmark$ &$\checkmark$&$\checkmark$& $\checkmark$&$\checkmark$& & \small{-B 40CM \,-B 50CM}&&&&\\
        \hline
        J1939+2134&$\checkmark$ &$\checkmark$&$\checkmark$ & $\checkmark$&$\checkmark$&4 & \small{-B 40CM \,-B 50CM}&&&&\\
        \hline
        J2124-3358&$\checkmark$ &$\checkmark$& & &$\checkmark$& & \small{-B 20CM}&&&&\\
        \hline
        J2129-5721&$\checkmark$ &$\checkmark$&& &$\checkmark$& & &&&&\\
        \hline
        J2145-0750&$\checkmark$ &$\checkmark$&$\checkmark$ & &$\checkmark$& & &\!\!\!\!\small{CPSR2\_50CM}&$\checkmark$&&\\
        \hline
        J2241-5236&$\checkmark$ &$\checkmark$&$\checkmark$ & &$\checkmark$& & &&&&\\
        \hline
    \end{tabular}
    \vspace{2ex}
    
    $^{\rm a}$ The figures in this column represent the chromaticity of the chromatic noise (CN). \\
    $^{\rm b}$ Shorthand notation for an exponential dip modeled by a exponential function. \\
    $^{\rm c}$ Shorthand notation for a Gaussian bump modeled by a Gaussian function. \\
    $^{\rm d}$ Shorthand notation for annual dispersion measure variation modeled by a yearly sinusoid. 
\end{table*}

In the timing residuals analysis, the cross power spectral density caused by the SGWB between any two pulsars, $a$ and $b$, takes 
\m
S_{ab}(f)=\Gm_{ab}\frac{h_c^2}{12\pi^2 f^3}=\Gm_{ab}\frac{A_{\GWB}^2}{12\pi^2}\left(\frac{f}{f_{\yr}}\right)^{-\gm}f_{\yr}^{-3},
\n
with $\gamma$ related to $\alpha$ as $\gamma=3-2\alpha$ and hence $\gamma=13/3$ for the TT mode, and $\gamma=5$ for all of the ST, VL, and SL modes. $\Gamma_{ab}$ is the overlap reduction function that describes the correlations between the pulsars and encodes the information about polarizations. The overlap reduction function can be calculated by \cite{Chamberlin:2011ev}:
\m
\Gm_{ab}^{P}=&&\frac{3}{8\pi} \int d \hat{\Om}(e^{2\pi i f L_a(1+\hat{\Om}\cdot \hat{p}_a)}-1) \times \notag \\
&&(e^{2\pi i f L_b(1+\hat{\Om}\cdot \hat{p}_b)}-1)F_a^P(\hat{\Om})F_b^P(\hat{\Om}),
\n
where the antenna patterns $F^P(\hat{\Om})$ are related to the polarization tensor $\epsilon_{ij}^P$ as
\m
F^P(\hat{\Om})=\frac{\hat{p}_i\hat{p}_j}{2(1+\hat{\Om}\cdot\hat{p})}\epsilon_{ij}^P(\hat{\Om}),
\n 
in which $P=+,\times$ denotes the two tensor modes, $P=x, y$ denotes the two vector modes, and $P=b, l$ denotes the scalar breathing and scalar longitudinal modes, respectively. The $+, \times$, and $b$ are transverse, while $x, y,$ and $l$ are longitudinal. The other physical quantities in the above definition are the distance between the Earth and pulsar $L$,  the propagation direction of the GW $\hat{\Om}$,  and the direction of the pulsar with respect to the Earth $\hat{p}$.
The TT, ST, VL, and SL correlations are defined by \cite{Cornish:2017oic}:
\m
&&\Gm_{ab}^{\rm{TT}}=\Gm_{ab}^{+}+\Gm_{ab}^{\times},\quad
\Gm_{ab}^{\rm{ST}}=\Gm_{ab}^{b},\notag\\
&&\Gm_{ab}^{\rm{VL}}=\Gm_{ab}^{x}+\Gm_{ab}^{y},\quad 
\Gm_{ab}^{\rm{SL}}=\Gm_{ab}^{l}.
\n
For the transverse modes, one has \citep{Lee,Chamberlin:2011ev,Qin:2018yhy}:
\m
\Gm_{ab}^{\rm{TT}}&=&\half\left[1+\dt_{ab}+3\kappa_{ab}\(\rm{ln}\,\kappa_{a b}-\frac{1}{6}\)\right],\\
\Gm_{ab}^{\rm{ST}}&=&\half\left(1+\dt_{ab}-\half\kappa_{ab}\right),	
\n
where $\kappa_{ab}=(1-\cos \xi_{ab})/2$, with $\xi_{ab}$ being the angle between the two pulsars. In the case of longitudinal polarization modes, the corresponding overlap reduction functions   $\Gm_{ab}^{\rm{VL}}$ and $ \Gm_{ab}^{\rm{SL}}$ have no analytical expressions, and are estimated numerically using the HCubature.jl package.\footnote{https://github.com/JuliaMath/HCubature.jl} $\Gm_{ab}^{\rm{TT}}$ is the well-known Helling $\&$ Downs correlation \citep{Hellings:1983fr} that is deemed to be the criterion for the detection of an SGWB predicted by general relativity. Similarly, the presence of $\Gm_{ab}^{\rm{ST}}$, $\Gm_{ab}^{\rm{VL}}$ or $\Gm_{ab}^{\rm{SL}}$ correlations indicates the detection of the SGWBs from a modified gravitational theory. 


\section{PTA data analysis}\label{sec3}

\begin{table*}[!htbp]
    \footnotesize
    \caption{Parameters and their prior distributions used in the analyses.}
    \label{prior}
    \begin{tabular}{c c c c}
        \hline
        Parameter & Description & Prior & Comments  \,\\
        \hline
        \multicolumn{4}{c}{White noise}\,\\	        
        $E_{k}$ & EFAC per backend/receiver system & $\uni[0, 10]$ & single pulsar analysis only \\
        $Q_{k}$[s] & EQUAD per backend/receiver system & $\logu[-8.5, -5]$ & single pulsar analysis only \\
        $J_{k}$[s] & ECORR per backend/receiver system & $\logu[-8.5, -5]$ & single pulsar analysis only \\
        \hline
        \multicolumn{4}{c}{Red noise (including SN, DM and CN)} \\
        $\Am_{\RN}$ & red noise power-law amplitude &$\logu[-20, -8]$ & one parameter per pulsar\, \\
        $\gamma_{\RN}$ &red noise power-law index  &$\uni[0,10]$ & one parameter per pulsar\, \\
        \hline
        \multicolumn{4}{c}{Band/System noise}\,\\
        $\Am_{\BN/\GN}$ & band/group-noise power-law amplitude &$\logu[-20, -8]$ & one parameter per band/system\, \\
        $\gamma_{\BN/\GN}$ &band/group-noise power-law index &$\uni[0,10]$ &one parameter per band/system\, \\
        \hline
        \multicolumn{4}{c}{Deterministic event}\,\\
        $\Am_{\mathrm{E}}$ & exponential dip amplitude &$\logu[-10, -2]$ & one parameter per exponential dip event \, \\
        $ t_{\mathrm{E}}[\mrm{MJD}]$ &time of the event & $\uni[57050, 57150]$ for PSR J1643 & one parameter per exponential dip event \, \\
        $ $ & &$\uni[56100, 56500]$ for PSR J2145 & \, \\
        $ $ & &$\uni[54500, 54900]$ for PSR J1713 $^{1}$& \, \\
        $ $ & &$\uni[57500, 57520]$ for PSR J1713 $^{2}$& \, \\
        $\mrm{log_{10}} \tau_{\mrm{E}}[\mrm{MJD}]$ & relaxation time for the dip &$\uni[\mrm{log_{10}}5, 2]$ &one parameter per exponential-dip event \, \\
        $\Am_{\mathrm{G}}$ & Gaussian bump amplitude &$\logu[-6, -1]$ & one parameter per Gaussian bump event \, \\
        $ t_{\mrm{G}}[\mrm{MJD}]$ & time of the bump &$\uni[53710, 54070]$ &one parameter per Gaussian bump event \, \\
        $ \sigma_{\mrm{G}}[\mrm{MJD}]$ & width of the bump&$\uni[20, 140]$ &one parameter per Gaussian bump event \, \\
        $\Am_{\mathrm{Y}}$ & annual variation amplitude &$\logu[-10, -2]$ & one parameter per annual event \, \\
        $ \phi_{\mrm{Y}}$ & phase of the annual variation &$\uni[0, 2\pi]$ &one parameter per annual event \, \\
        \hline
        \multicolumn{4}{c}{Common-spectrum Process}\,\\
        $\Am_{\mrm{UCP}}$ & UCP power-law amplitude &$\logu[-18, -14]$ & one parameter per PTA\, \\
        $\Am_{\mrm{TT}}$ & SGWB amplitude of TT polarization &$\logu[-18, -14]$ & one parameter per PTA\, \\
        $\Am_{\mrm{ST}}$ & SGWB amplitude of ST polarization &$\logu[-18, -14]$ & one parameter per PTA\, \\
        $\Am_{\mrm{VL}}$ & SGWB amplitude of VL polarization &$\logu[-19, -15]$ & one parameter per PTA\, \\
        $\Am_{\mrm{SL}}$ & SGWB amplitude of SL polarization &$\logu[-20, -16]$ & one parameter per PTA\, \\
        \hline
    \end{tabular}
    \vspace{2ex}
    
    $^{1,}$$^{2}$ There are two exponential dip events for pulsar J1713+0437.   
\end{table*}

To effectively extract the spatially correlated signals, one needs to provide a possibly comprehensive description of the arrival time variations induced by various stochastic effects. In the analyses, we adopt the noise model developed in \cite{Goncharov:2020krd} with possible deterministic and stochastic processes. After subtracting the timing model of the pulsar from the TOAs, the timing residuals $\dt \bt$ can be decomposed into
\m
\dt \bt = M \be +F \ba + \bd +\bn.
\label{nm}
\n

The first term $M \be$ is the linear term when Taylor-expanding the timing model around the estimated parameters, where $M$ is the design matrix and $\be$ is the vector of the offset parameters, i.e., the difference between the true parameters and the estimated parameters. 

The second term $F\ba$ represents stochastic red noise, where $F$ is the Fourier design matrix that incorporates a radio frequency-dependent term $\bnu_{i}^{-\chi}$ and alternative sine and cosine components at frequencies  $\{1/T, 2/T, ...N_{\mathrm{mode}}/T\}$,  with $\bnu_{i}$ being the radio frequency of the $i$-th TOA, $\chi$ the chromaticity of the noise, $T$ the time span of the observation, $N_{\mathrm{mode}}$ the number of Fourier frequencies used, and $\ba$ is the vector of alternating sine and cosine amplitudes. The red noises come from several sources. For example, the irregular motion of the pulsar itself contributes an achromatic red noise called spin noise (SN) with $\chi=0$ \citep{Shannon:2010bv}; the change in column density of the ionized plasma in the interstellar medium causes frequency-dependent dispersion measure (DM) variations with $\chi=2$ \citep{Keith:2012ht}; and scattering variations in interstellar medium lead to chromatic noise (CN) with $\chi=4$ \citep{Andrew:cn}. Moreover, band noise (BN) and system (``group") noise (GN) are separate red noise processes in a given band or system,  which may be produced by instrumental artifacts or interstellar processes that are incoherent between bands \citep{Lentati:2016ygu}. Following \cite{Arzoumanian:2020vkk}, we use 30 frequency components ($N_{\mathrm{mode}} = 30$) for the red noise of the individual pulsar, and use 5 frequency components ($N_{\mathrm{mode}} = 5$) for the common process among all of the pulsars. 

The third term $\bd$ represents deterministic signals, including chromatic exponential dips, extreme scattering events, annual dispersion measure variations, and system-dependent profile evolution \citep{Goncharov:2020krd}.
The first two situations can be attributed to the sudden change in dispersion or scattering when the signal passes through the interstellar medium during propagation \citep{Lentati:2016ygu,Keith:2012ht}, and the annual DM variations are manifested as the results of gradient changes in electron column density between the pulsar and the Earth, caused by the motion of the Earth around the Sun \citep{Coles:2015uia}; the three kinds of events are respectively described by an exponential function, a Gaussian function, and a yearly sinusoid, respectively.
The system-dependent profile evolution is characterized as a linear function of the frequency of some systems in order to help model the pulsar J0437-4715 \citep{Goncharov:2020krd}. 

The last term $\bn$ represents the white noise that is modeled by the TOA uncertainties and the three parameters EFAC, EQUAD, and ECORR \citep{NANOGrav:2015qfw}. Specifically, EFAC scales the TOA uncertainty, EQUAD adds an extra term independent of uncertainty, and ECORR describes the excess variance for sub-banded observations.


In the noise analyses, the Bayesian inference has been used in determining the noises existing in the TOAs of a certain pulsar \citep{Goncharov:2020krd}, and we also use the method to decide the preferred model from the possible candidates in our analyses, by calculating the Bayes factor (BF). To be specific, given the observed data set $\Dm$, for two models  $\Hm_0$ and $\Hm_1$,  the BF is 
\m
\rm{BF}=\frac{\rm{Pr}(\Dm|\Hm_1)}{\rm{Pr}(\Dm|\Hm_0)}, 
\n 
where $\rm{Pr}(\Dm|\Hm)$ is the evidence that measures the probability of obtaining the data $\Dm$ under the hypothesis of model $\Hm$. Usually, only when $\rm{BF}>3$ can one declare a positive preference for $\Hm_1$ over $\Hm_0$ \citep{BF}. In practice, we use the \textit{product-space method} \citep{10.2307/2346151,10.2307/1391010,Hee:2015eba,Taylor:2020zpk} to estimate the BFs, as was done in \cite{Arzoumanian:2020vkk}.

In this work, we search for the nontensorial SGWBs in PPTA DR2 by excluding the pulsar J0437$-$4715 because it is challenging to obtain a complete noise model for this pulsar \citep{Goncharov:2020krd}. The noise models for the 25 pulsars used in the analyses are listed in \Table{noise}. We use the recent DE438 \citep{DE438} as the fiducial solar system ephemeris, and fix the white noise parameters to their maximum likelihood values from the single pulsar noise analysis based on the noise model shown in \Table{noise}. We perform parameter estimations using the PTMCMCSampler package \citep{justin_ellis_2017_1037579} with the likelihood and BF being evaluated with the enterprise \citep{enterprise} and enterprise\_extension \citep{enterprise_extensioins} packages. All of the parameters and their prior distributions are listed in \Table{prior}.  
\section{results and discussion}\label{sec4}

The PPTA collaboration found significant evidence for a common-spectrum process in their DR2 data set, with the BF of a spatially uncorrelated common-spectrum process (UCP) versus the null model with no common-spectrum process being larger than $3\times 10^{6}$ \citep{Goncharov:2021oub}. However, the posterior of the spectral slope $\gamma_{\mrm{UCP}}$ has a rather broad distribution, indicating we are not able to distinguish the different astrophysical processes that can result in the different spectral slopes of the SGWB.
 
\begin{table}[hb]
    \centering
	\begin{tabular}{c| c c c c}
		\hline
		\hline
		Model & TT & ST  & VL & SL \\
		\hline
		BF & 2.15(4) &0.183(3) & 1.06(2) & 0.362(6) \\
		\hline
	\end{tabular}
	\caption{BFs of the TT, ST, VL, and SL models compared to the UCP model. The digit in the parentheses gives the uncertainty on the last quoted digit.}
	\label{BF}
\end{table}  
   
\begin{figure}[htbp!]
    \centering
    \includegraphics[width = 0.48\textwidth]{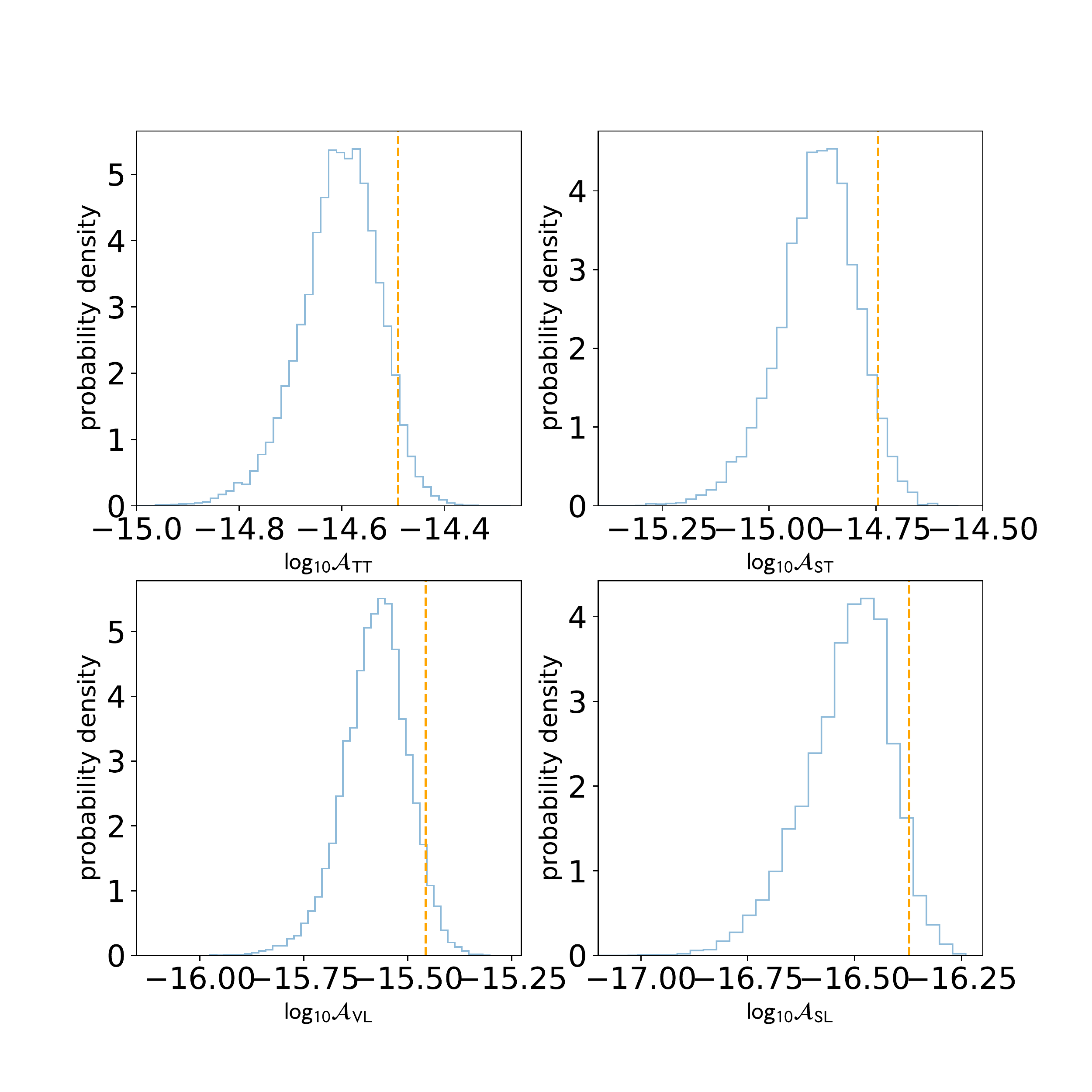}\caption{ \label{gw_lgA_pd}The marginalized posteriors for the amplitudes obtained from the TT, ST, VL, and SL models, respectively. The orange dashed vertical lines lie at $95\%$ credible intervals of the corresponding amplitudes.}
\end{figure}

We use the UCP as the fiducial model and report the BFs of different models with respect to the UCP model in \Table{BF}. The BFs for all of the TT, ST, VL, and SL models are smaller than 3, implying no statistically significant Bayesian evidence for an SGWB with the TT, ST, VL, or SL spatial correlations in PPTA DR2. We therefore place the upper limits for the amplitudes and their posterior distributions as shown in \Fig{gw_lgA_pd}. The $95\%$ upper limits for the amplitudes are $\mathcal{A}_{\mrm{TT}} \lesssim 3.2\times 10^{-15}$, $\mathcal{A}_{\mrm{ST}} \lesssim 1.8\times 10^{-15}$, $\mathcal{A}_{\mrm{VL}} \lesssim 3.5\times 10^{-16}$, and $\mathcal{A}_{\mrm{SL}} \lesssim 4.2\times 10^{-17}$; or, equivalently, the $95\%$ upper limits for the energy density parameter per logarithm frequency are $\Omega_{\mathrm{GW}}^{\mathrm{TT}} \lesssim 1.4\times 10^{-8}$, $\Omega_{\mathrm{GW}}^{\mathrm{ST}} \lesssim 4.5\times 10^{-9}$, $\Omega_{\mathrm{GW}}^{\mathrm{VL}} \lesssim 1.7\times 10^{-10}$, and $\Omega_{\mathrm{GW}}^{\mathrm{SL}} \lesssim 2.4\times 10^{-12}$, at a frequency of 1/yr.

   
\begin{figure}[htbp!]
	\centering
	\includegraphics[width = 0.48\textwidth]{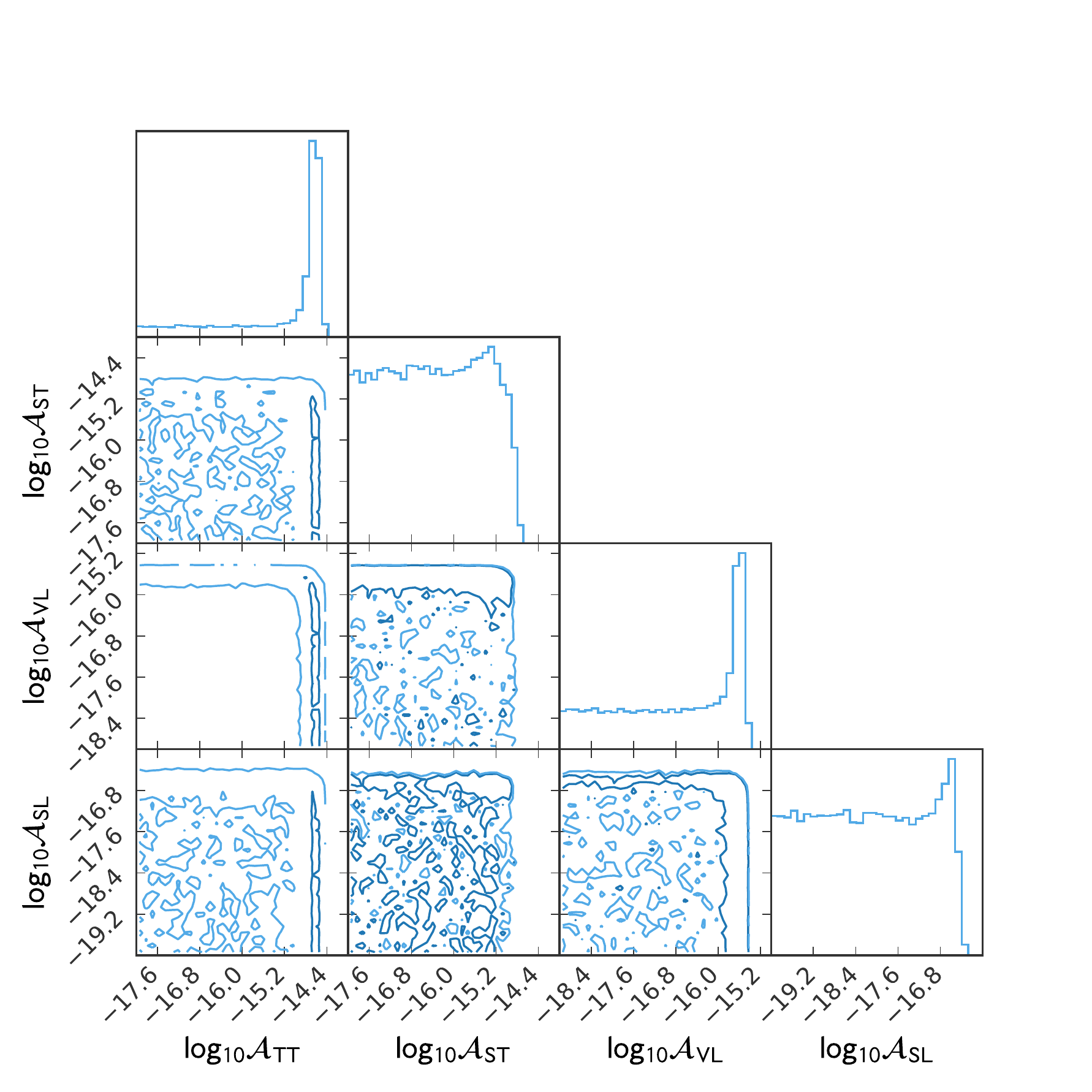}\caption{ \label{gw_lgA}One- and two-dimensional marginalized posteriors for the amplitudes obtained from the TT+ST+VL+SL model. We show both the $1\sigma$ and $2\sigma$ contours in the two-dimensional plots.}
\end{figure}


We further consider a TT+ST+VL+SL model, in which we simultaneously take all of the possible correlations into account. The posterior distributions for the amplitudes resulting from this model are shown in \Fig{gw_lgA}. 
The corresponding $95\%$ upper limits for the amplitudes are $\mathcal{A}_{\mrm{TT}} \lesssim 3.0\times 10^{-15}$, $\mathcal{A}_{\mrm{ST}} \lesssim 1.0\times 10^{-15}$, $\mathcal{A}_{\mrm{VL}} \lesssim 3.0\times 10^{-16}$, and  $\mathcal{A}_{\mrm{SL}} \lesssim 2.7\times 10^{-17}$; or, equivalently, the $95\%$ upper limits for the energy density parameter per logarithm frequency are $\Omega_{\mathrm{GW}}^{\mathrm{TT}} \lesssim 1.2\times 10^{-8}$, $\Omega_{\mathrm{GW}}^{\mathrm{ST}} \lesssim 1.4\times 10^{-9}$,  $\Ovlgw \lesssim 1.2\times 10^{-10}$ and $\Oslgw \lesssim 1.0\times 10^{-12}$, at a frequency of 1/yr. 
With four signals competing in the TT+ST+VL+SL model, the corresponding amplitudes of each polarization are smaller than those in the model using only one polarization.

The BF of the ST model versus the UCP model is $0.183\pm 0.003$, implying the PPTA DR2 shows no significant Bayesian evidence for (or against) the ST spatial correlations in the data. However, the upper limit for the amplitude of ST polarization constrained by PPTA DR2 is still consistent with the results reported in \cite{Chen:2021wdo}, where $\Omega_{\mathrm{GW}}^{\mathrm{ST}}=1.54^{+1.20}_{-0.71}\times 10^{-9}$. Therefore, the physical origin of the ST process reported in \cite{Chen:2021wdo} remains to be answered by the future PTA data sets with increasing time spans and numbers of pulsars.

\begin{acknowledgments}
We thank the anonymous referee for the useful suggestions and comments. We also acknowledge the use of the HPC Cluster of ITP-CAS and the HPC Cluster of Tianhe II in the National Supercomputing Center in Guangzhou. This work is supported by the National Key Research and Development Program of China, Grant No.2020YFC2201502, grants from NSFC (grants No. 11975019, 11690021, 11991052, and 12047503),  the Key Research Program of the Chinese Academy of Sciences (grant No. XDPB15), the Key Research Program of Frontier Sciences, CAS, grant No. ZDBS-LY-7009, the science research grants from the China Manned Space Project with No. CMS-CSST-2021-B01, and CAS Project for Young Scientists in Basic Research with No. YSBR-006 .
\end{acknowledgments}
\bibliography{./bibfile-ppta}
\bibliographystyle{aasjournal}

\end{document}